\documentclass[12pt,aps,groupedaddress,reprint,showpacs,nofootinbib,prd]{revtex4-1}
\usepackage{amsmath,amssymb,mathrsfs,graphicx,slashed,ragged2e,caption}

\usepackage{hyperref}
\usepackage{graphicx}
\usepackage{dcolumn}
\usepackage{bm}
\usepackage{caption}

\DeclareMathOperator{\Tr}{Tr}
\newcommand{\mM}{\text{Met}(M)}

\begin{document}

\title[The configuration manifold of neutron star models]{A metric on the space of neutron star models in general relativity and modified gravity}

\author{Arthur George Suvorov}
\email{arthur.suvorov@tat.uni-tuebingen.de}
\address{Theoretical Astrophysics, IAAT, University of T{\"u}bingen, Germany}

\date{\today}

\begin{abstract}
Some pairs of neutron star models can intuitively be thought of as being `closer' together than others, in the sense that more precise observations might be required to distinguish between them than would be necessary for other pairs. In this paper, borrowing ideas from the study of geometrodynamics, we introduce a mathematical formalism to define a geometric distance between stellar models, to provide a quantitative meaning to this notion of closeness. In particular, it is known that the set of all metrics on a Riemannian manifold itself admits the structure of a Riemannian manifold (`configuration manifold'), which comes equipped with a canonical metric. By thinking of a stationary star as being a particular $3+1$ metric, the structure of which is determined through the Tolman-Oppenheimer-Volkoff relations and their generalisations, points on a suitably restricted configuration manifold can be thought of as representing different stars, and distances between these points can be computed. We develop the necessary mathematical machinery to build the configuration manifold of neutron star models, and provide some worked examples to illustrate how this space might be used in future studies of stellar structure.
\end{abstract}


\maketitle

\section{Introduction}

A fundamental problem in high-energy physics concerns the behaviour of macroscopic matter at nuclear densities. A determination of the nuclear equation of state (EOS) of matter within neutron stars (NSs) provides an astrophysical avenue for studying this problem. Indeed, given a relationship between the thermodynamic variables of the star, such as the pressure, density, and temperature, a unique stellar model can be constructed \cite{shapteu}. Owing to the complexity of the physical processes involved, many different EOS have been proposed \cite{nseos1,nseos2}, each of which predict different macroscopic properties for the stars. Modified gravity considerations also complicate the picture, since the compactness of the star, and hence the mass-radius relationship, is ultimately determined by the ability for the hydrostatic pressure to resist gravitational compression \cite{modgrns1,modgrns2}. Observations of NS masses and radii, coming from analyses of their oscillation \cite{pulsate}, gravitational wave \cite{gws}, and X-ray \cite{xrays} spectra, can be used to identify the EOS by solving the respective inverse problems \cite{inverse1,inverse2,inverse3}.

Suppose that the configuration space of viable NS models (or some subset thereof), i.e. the set of EOS consistent with causal constraints \cite{ruffini} and observations, can be built (such as the set considered in \cite{most18}). There should be a way to think about different members of this space as being `close' to one another (or otherwise), in the sense that elements which make quantitatively similar predictions should neighbour each other. It is the purpose of this paper to propose a formal, mathematical means to achieve this, based on geometrodynamic concepts first introduced by Wheeler \cite{wheeler1,wheeler2}. In general, an EOS corresponds to a spacetime metric through the Tolman-Oppenheimer-Volkoff relations or their generalisations \cite{gentov}. This allows one to think about the space of stars defined by a set of EOS as being equivalent to a specific collection of metrics.

Given a manifold $M$, it is known that the collection of all Riemannian metrics on $M$, $\mM$, itself admits the structure of an infinite-dimensional Riemannian manifold \cite{gil1,gil2,clark10,demmel}. As such, given two metrics $h$ and $k$ on $M$, the metric $G$ on $\mM$ essentially defines an inner product between tangent vectors at the `points' $h, k \in \mM$. In the context of general relativity, Wheeler called this collection \emph{superspace} and used it to study the configuration space of cosmology \cite{wheeler1,wheeler2} and the concept of quantum foam \cite{wheeler3}. The structure of Wheeler's superspace has since been investigated in detail \cite{fishcer,edwards,guil09}, though without much attention towards its potential application to stellar dynamics.


Here, we restrict our attention to only those metrics $h$ and $k$ which correspond to NS geometries. This allows us to consider a finite-dimensional submanifold $\text{Met}_{\text{NS}}(M) \subset \mM$  (cf. \cite{sens1,sens2}), whose geometric properties can be tied to stellar physics in a precise way. In particular, this submanifold is parameterised by the macroscopic stellar variables, such as the masses and radii of the stars, rather than the usual spacetime coordinates. This allows for a natural means to quantify the relationship between different EOS. The formalism has the benefit that one is not restricted to general relativity or any other particular theory of gravity \emph{a priori}, since the only inputs are the actual metrics themselves. Typically, stars are defined within the context of a \emph{spacetime} $\mathcal{M}$, though we can perform a $3+1$ split to work with a spacelike hypersurface $M \subset \mathcal{M}$, which is explicitly Riemannian. Having constructed a metric $G$ on $\text{Met}_{\text{NS}}(M)$, we can further determine the Christoffel symbols, and thus geodesic curves, from which explicit distances can be computed.

This short paper is organised as follows. In Section 2 we introduce the mathematical machinery surrounding the configuration manifold. Section 3 explores the relevance of this space to NS geometries, and in Section 4 we evaluate the metric (4.1) and compute geodesics (4.2) for a specific case of Tolman VII stars to provide a worked example. Some discussion is offered in Section 5.

\section{The configuration manifold}

As mentioned in the Introduction, the set of all Riemannian metrics over a Riemannian manifold $M$ admits the structure of an infinite-dimensional Riemannian manifold \cite{gil1,gil2}, denoted $\mM$. In this sense, points of $\mM$ are Riemannian metrics on $M$: each $p \in \mM$ corresponds to a positive-definite, symmetric $(0,2)$-tensor over $M$. We consider $M$ to be 3-dimensional (though a generalisation to higher dimensions is straightforward), as later it will be identified with a spacelike hypersurface, defined via a $3+1$ splitting, of a NS spacetime.

If the manifold $M$ is compact, then one may introduce a metric\footnote{Some other choices are possible under certain conditions, see \cite{clarkphd}. The metric \eqref{eq:gilmed} used here is called the \emph{canonical metric}, as it is invariant under the action of the diffeomorphism group $\text{Diff}(M)$ on $\mM$. Note also that the compactness of $M$ is sufficient but not necessary; see Sec. 5.}, in the $L^{2}$-topology \cite{clark10}, over $\mM$ as \cite{gil1,gil2,demmel}
\begin{equation} \label{eq:gilmed}
G(\alpha,\beta) = \int_M d^{3} x \sqrt{g} \Tr \left( g^{-1} \alpha g^{-1} \beta \right) ,
\end{equation}
where $\alpha$ and $\beta$ are tangent vectors to the space of metrics at the `point' $g$, which serves as a reference metric. On an intuitive level, a measure of distance between two vectors naturally depends on the choice of basis and the origin, which is why it is necessary to introduce a base metric $g$ within \eqref{eq:gilmed}. 

As it stands, the metric \eqref{eq:gilmed} is defined over the infinite-dimensional manifold $\mM$, which is difficult to work with. We restrict the domain to one in which $\alpha$ and $\beta$ only correspond to tangent vectors to the space of \emph{NS metrics} (see Sec. 3). That is, we consider a submanifold $\text{Met}_{\text{NS}}(M) \subset \mM$, which inherits a metric, which we also call $G$ through a slight abuse of notation, from its parent space via pullback (cf. \cite{sens1,sens2}). It is difficult to define the submanifold $\text{Met}_{\text{NS}}(M)$ in total generality since, depending on the included physics, there may be an arbitrarily large (but finite) number of parameters which describe the stellar model; the stress-energy tensor may be arbitrarily complicated. Nevertheless, suppose that a star can be described by $N$ macroscopic parameters: $q^{1}, \ldots, q^{N}$, e.g. mass, radius, central temperature, polar magnetic field strength, rotational frequency, and so on. These parameters $\boldsymbol{q}$ define a natural coordinate basis for the $N$-dimensional space $\text{Met}_{\text{NS}}(M)$ (see Sec. 4). 

With respect to this basis, the tensor components of \eqref{eq:gilmed} read \cite{sens1,suv19}
\begin{equation} \label{eq:metten}
G_{ij} = \int_{M} d^{3} x \sqrt{g}  g^{nk} \frac { \partial g_{mn}} {\partial q^{i}}   g^{\ell m}  \frac { \partial g_{\ell k}} {\partial q^{j}}  ,
\end{equation}
where $1 \leq i,j \leq N$. From \eqref{eq:metten}, the relevant geometric quantities of $\text{Met}_{\text{NS}}(M)$ can be defined, including the Christoffel symbols $\Gamma$. The distance between two metrics $h$ and $k$, described by parameters $\boldsymbol{q}_{h}$ and $\boldsymbol{q}_{k}$, respectively, is then given by the length of a geodesic\footnote{Note that, while the space $\text{Met}(M)$ is \emph{not} geodesically complete (see Corollary 2.47 of \cite{clarkphd}), if the domains of the parameters $\boldsymbol{q}$ are finite, then the submanifold $\text{Met}_{\text{NS}}(M)$ will be complete as a consequence of the Hopf-Rinow theorem. This ensures that the geodesic connecting any two `stars' will be well-defined.} $\gamma: [a,b] \mapsto \text{Met}_{\text{NS}}(M)$ connecting these points, viz.
\begin{equation} \label{eq:distance}
d(h,k) = \int^{b}_{a} d \tau \sqrt{ G_{ij} \frac {d \gamma^{i}} {d \tau} \frac {d \gamma^{j}} {d \tau} },
\end{equation}
for affine parameter $\tau$, where $\gamma^{j}(a) = q_{h}^{j}$ and $\gamma^{j}(b) = q_{k}^{j}$, and $\boldsymbol{\gamma}$ satisfies the geodesic equation,
\begin{equation}
0 = \frac {d^2 \gamma^{i}} {d \tau^2} + \Gamma^{i}_{j k} \frac {d \gamma^{j}} {d \tau} \frac {d \gamma^{k}} {d \tau}.
\end{equation}


\section{Neutron star geometries}

In general relativity and other theories of gravity, one typically deals with a \emph{spacetime} $(\mathcal{M},\kappa)$, which is Lorentzian and not Riemannian, i.e. the metric $\kappa$ is not positive-definite. The space $\mM$ described above therefore cannot be constructed immediately from a given set of stellar models. One must first extract a Riemannian manifold from the 4-dimensional spacetime. If the spacetime is stationary, a unique extraction can be achieved through a $3+1$ split \cite{wald84}.

As such, we restrict our attention to NS spacetimes $(\mathcal{M},\kappa)$ which are stationary, so that there exists a timelike Killing vector $\boldsymbol\xi$ satisfying \cite{wald84} 
\begin{equation} \label{eq:killing}
\nabla_{\mu} \xi_{\nu} + \nabla_{\nu} \xi_{\mu} = 0 .
\end{equation}
One may now define the norm, $\lambda$, and twist, $\boldsymbol{\omega}$, of $\boldsymbol\xi$ through
\begin{equation} \label{eq:norm}
\lambda = \xi^{\alpha} \xi_{\alpha} ,
\end{equation}
and 
\begin{equation} \label{eq:twist}
\omega_{\alpha} = \epsilon_{\alpha \beta \gamma \delta} \xi^{\beta} \nabla^{\gamma} \xi^{\delta} ,
\end{equation}
respectively. A general line element on $\mathcal{M}$ may now be written in the generalised Papapetrou form \cite{hansen,simbei83}
\begin{equation} \label{eq:papapet}
\begin{aligned}
ds^2 &= \kappa_{\mu \nu} dx^{\mu} dx^{\nu} \\
&= \lambda ( dt + \sigma_{i} dx^{i} )^{2} - \lambda^{-1} h_{ij} dx^{i} dx^{j} ,
\end{aligned}
\end{equation}
where the twist $\boldsymbol{\omega}$ is related to $\boldsymbol{\sigma}$ through $\omega_{i} = - \lambda^{-2} \epsilon_{ijk} D^j \sigma^{k}$ and $D$ forms the covariant derivative with respect to $h$. The form \eqref{eq:papapet} illustrates a $3+1$ split of the spacetime $(\mathcal{M},\kappa)$, and we denote the manifold associated with the Riemannian 3-metric $h$ as $S$. It is this class of metrics $h$ that form the inputs for the metric $G$ on the configuration manifold \eqref{eq:metten}, once a suitable restriction of $S$ is considered. {Note that in writing \eqref{eq:papapet} we have chosen the time coordinate $t$ such that $\xi^{\mu} = (1,0,0,0)$, which ensures that the (time-independent) potentials $\lambda$, $\sigma_{i}$, and $h_{ij}$ are uniquely defined given some stationary $\kappa_{\mu \nu}$.
}

Indeed, recall that we considered only compact manifolds $M$ to ensure that the integral within \eqref{eq:gilmed} converges. However, since we wish to measure the difference between two stellar configurations, it is reasonable to consider only the section of $S$ confined by {some notion of the maximal} stellar surface, i.e. we consider $M \subset S$, where $M$ is defined by the presence of a {(maximal)} non-zero stress energy tensor (see below). This $M$ is to be identified with the domain of the integral \eqref{eq:gilmed}. However, in general, two stars will define {different spheroids}, and so care must be taken to ensure that the whole star is always considered. 

{For any collection of NS models, the space $M \subset S$ is built to include all possible stars within the family, i.e. it is defined as the union of points \emph{potentially} occupied by matter. For spherical stars, this amounts to identifying the largest stellar radius permitted by the EOS (see Sec. 4). However, for rotating stars, which are often not contained within one another in a concentric sense, one must identify $M$ with the set of permitted spheroids, i.e. the volume defined by the union of points defining the maximally oblate and prolate stars permitted by the EOS (set, for instance, by the centrifugal breakup limit).}

In general, the components of $h$ defined within \eqref{eq:papapet} are to be subjected to some set of field equations. One typically introduces a stress-energy tensor $\boldsymbol{T}$, which is non-zero inside the star, though vanishes outside, which acts as a source for $\kappa$ and hence $h$. {For example, neglecting viscosity, anisotropic pressures, and magnetic fields, a simple NS may be composed of a perfect-fluid \cite{shapteu} with stress-energy}
\begin{equation} \label{eq:perffluid}
T^{\mu \nu} = \left( \rho + p \right) u^{\mu} u^{\nu} + p \kappa^{\mu \nu},
\end{equation}
{for mass density $\rho$, pressure $p$, and 4-velocity $\boldsymbol{u}$, where we have taken natural units $c=G=1$. In general relativity, the Einstein equations}
\begin{equation}
R_{\mu \nu} - \frac {1} {2} R^{\alpha}_{\alpha} \kappa_{\mu \nu} = 8 \pi T_{\mu \nu},
\end{equation}
{where $R_{\mu \nu}$ is the Ricci tensor, determine the structure of $\boldsymbol{\kappa}$. The union of points for which $p$ does not vanish thus defines the set $M$. In a modified theory of gravity, the metric $\boldsymbol{\kappa}$ (and hence $\boldsymbol{h}$) is similarly determined through a modified set of field equations. In an $f(R)$ theory of gravity, for instance, the metric is set through the $f(R)$ field equations (e.g. \cite{suvmel})
\begin{equation}
\begin{aligned}
&f'(R^{\alpha}_{\alpha}) R_{\mu \nu} -  \frac {f(R^{\alpha}_{\alpha})} {2} \kappa_{\mu \nu} \\
&+ \left( \kappa_{\mu \nu} \square - \nabla_{\mu} \nabla_{\nu} \right) f'(R^{\alpha}_{\alpha}) = 8 \pi T_{\mu \nu}.
\end{aligned}
\end{equation}
}

The components $h_{ij}$ satisfy boundary conditions across the stellar surface, defined as the vanishing of the stress-energy \eqref{eq:perffluid}, so as to continuously match the geometry to some exterior \cite{israel}. For instance, static, spherically symmetric spacetimes in general relativity must match to an exterior Schwarzschild geometry by virtue of Birkhoff's theorem \cite{shapteu,wald84}. 

\subsection{Spherically symmetric stars}

To make the above more explicit, we consider the case of spherically symmetric stars, so that the various steps involved are clearly laid out. The general spacetime metric $\kappa$, in Boyer-Lindquist coordinates $(t,r,\theta,\phi)$ \cite{wald84}, is given by
\begin{equation} \label{eq:statmet}
ds^2_{\kappa} = -A(r) dt^2 + B(r) dr^2 + r^2 d \Omega^2.
\end{equation}
From \eqref{eq:papapet}, the line element on $S$ reads (e.g. \cite{suvmel})
\begin{equation} \label{eq:3dimmet}
ds^2_{S} = B(r) A(r) dr^2 + r^2 A(r) d \Omega^2.
\end{equation}

Consider any two stars, characterised by two distinct metrics of the form \eqref{eq:3dimmet}, where the first star has radius $R_{1}$, and the second has radius $R_{2}$. Without loss of generality, assume that $R_{2}  \geq R_{1}$. In the region $R_{1} \leq r \leq R_{2}$, the first spacetime is Schwarzschild, i.e. we have that
\begin{equation} \label{eq:schwarzschild}
A_{1}(r>R_{1}) = B_{1}(r>R_{1})^{-1} = 1 - \frac{2 M_{1}}{r}.
\end{equation}

To make sure that we capture the features of all stars within the set of models under consideration, it is important that the spacelike hypersurface $M$ is defined with respect to the largest radius within this set, i.e. the largest such $R_{2}$, which we call $\bar{R}$. {In general, $\bar{R}$ is defined as the largest radius for which a hydrostatic equilibrium exists for the EOS under consideration.} Suppose now that each member of the family within \eqref{eq:papapet} depends on some (maximal) set of parameters $\boldsymbol{q}$. In this case, the Riemannian manifold $\text{Met}_{\text{NS}}(M)$ is $\dim(\boldsymbol{q})$-dimensional, and the metric tensor \eqref{eq:metten} has components

\begin{equation} \label{eq:genmetten}
\begin{aligned}
G_{ij} =& 4 \pi \int^{\bar{R}}_{0} dr \frac {r^2} {\sqrt{ A B^3}} \Big[ A \frac {\partial B} {\partial q^{i}} \left( B \frac {\partial A} {\partial q^{j}} + A \frac {\partial B} {\partial q^{j}} \right) \\
&+ B \frac {\partial A} {\partial q^{i}} \left( 3 B \frac {\partial A} {\partial q^{j}} + A \frac {\partial B} {\partial q^{j}} \right) \Big], 
\end{aligned}
\end{equation}

where care is to be taken with regards to integration in the region $R \leq r \leq \bar{R}$.




\section{Worked Example: Tolman VII stars}

We consider a simple, worked example to demonstrate the mathematical machinery developed in the previous sections. As is well-known, the Tolman VII solution is an exact solution to the Einstein field equations {with perfect fluid matter} \eqref{eq:perffluid} \cite{tol39}. The advantage of this solution is that the stellar density $\rho$ has the simple form
\begin{equation} \label{eq:tolmandensity}
\rho(r) = \frac {15 M} {8 \pi R^3} \left[ 1 - \left( \frac {r} {R} \right)^{2} \right],
\end{equation}
for mass $M$ and radius $R$. Despite its simplicity, calculations of the binding energy and moment of inertia for NSs with more realistic EOS match well with those of the Tolman VII solution for $M \gtrsim M_{\odot}$ \cite{nseos1} (see also \cite{rag17}). {The Tolman VII density profile \eqref{eq:tolmandensity} has further been used to study compactness limits in scalar-tensor theories of gravity \cite{sot18}, and gravitational radiation from magnetically deformed \cite{magdef1,magdef2} and pulsating \cite{tolmodes} NSs in general relativity.}

A curious feature of the Tolman VII solution is that the stars exhibit no mass-radius relationship; both $M$ and $R$ are free parameters\footnote{Note, however, that one requires the compactness parameter $M G/ (R c^2) \lesssim 0.27$ to preserve causality, i.e. to ensure that the speed of sound is bounded by the speed of light \cite{ruffini}. {This sets a rather high value to the density maximum; $\rho_{\text{max}} = 6 \times 10^{18} \text{ kg m}^{-3}$ for a very compact star with radius of $R = 6 \times 10^{3} \text{ m}$.}}. This will not be the case for more realistic EOS, and other parameters, such as the central temperature, will feature instead.

\subsection{Metric functions}

In natural units, the metric functions $A$ and $B$ within \eqref{eq:statmet}, for the Tolman VII metric, read \cite{nseos1,tol39}
\begin{equation} \label{eq:tolman1}
A(r) = \left(1 - \frac {5 M} {3 R} \right) \cos^2 \left[ \Phi(r) \right],
\end{equation}
and
\begin{equation}
B(r) = \left[ 1 - \frac {M r^2} {R^3} \left(5 - \frac {3 r^2} {R^2} \right) \right]^{-1},
\end{equation}
where 
\begin{equation} \label{eq:tolman3}
\begin{aligned}
\Phi(r) =& \frac {1} {2} \log \left[ \frac {1 + 2 \sqrt{ \frac {3 R} {M} - 6}} {\frac {6 r^2} {R^2} - 5 + 2  \sqrt{ \frac {9 r^4} {R^4} - \frac {15 r^2} {R^2} + \frac {3R} {M} } } \right] \\
&+ \arctan \left[ \frac {M} {\sqrt{3 M \left( R - 2 M \right) }} \right].
\end{aligned}
\end{equation}
Outside of the star, $r > R$, the metric functions continuously match to the Schwarzschild exterior \eqref{eq:schwarzschild}. 

Given expressions \eqref{eq:tolman1}--\eqref{eq:tolman3}, one may evaluate the metric components \eqref{eq:genmetten}, to in turn measure the `distance' \eqref{eq:distance} between two Tolman VII configurations, one described by the pair $(R_{1},M_{1})$ and the other by $(R_{2},M_{2})$. The metric $G$ is therefore parameterised by the coordinates $(R,M)$ and we have that, for example,
\begin{equation} \label{eq:metcomp}
\begin{aligned}
G_{MM} =& 4 \pi \int^{\bar{R}}_{0} dr \frac {r^2} {\sqrt{ A B^3}} \Big[ A \frac {\partial B} {\partial M} \left( B \frac {\partial A} {\partial M} + A \frac {\partial B} {\partial M} \right) \\
&+ B \frac {\partial A} {\partial M} \left( 3 B \frac {\partial A} {\partial M} + A \frac {\partial B} {\partial M} \right) \Big]
\end{aligned}
\end{equation}

is the ``$M M$'' component of the metric tensor \eqref{eq:metten}. Actually evaluating the integral within \eqref{eq:metcomp} is, unfortunately, non-trivial owing to the logarithmic and trigonometric functions appearing within the functions $A$ and $B$ above, though can be evaluated numerically without much difficulty.

Table \ref{tab:simdata} shows distances $d(R_{1},M_{1},R_{2},M_{2})$ from \eqref{eq:distance} between distinct Tolman VII configurations for various stellar radii and masses. We see that, for fixed radius $R_{1} = R_{2}$, even for rather large variations in mass $1.2 \leq M / M_{\odot} \leq 2.0$, the distances are relatively small; $d \lesssim 10^{5}$ for $M_{2}/M_{1} \lesssim 1.4$ . In contrast, even for $\lesssim 5\%$ changes in the radius, the distance is relatively large for fixed mass $M_{1} = M_{2}$; $d \gtrsim 5 \times 10^{5}$ for $R_{1}/R_{2} \lesssim 0.95$. This shows that two configurations with the same radii but different masses are `closer together' than two configurations with the same masses but different radii. This is expected, since the central density within \eqref{eq:tolmandensity} varies strongly with radius, $\rho_{c} \propto R^{-3}$, while $\rho_{c}$ only varies linearly with $M$. Nevertheless, the mathematical framework captures this feature automatically.

\begin{table*}
\caption{Distances $d(R_{1},M_{1},R_{2},M_{2})$, defined in \eqref{eq:distance}, between various Tolman VII configurations \eqref{eq:tolman1}--\eqref{eq:tolman3}.}
\centering
  \begin{tabular}{llcccc}
  \hline
  \hline
$R_{1}$  ($10^{4} \text{ m}$) & $M_{1}$ ($M_{\odot}$)  & $R_{2}$ ($10^{4} \text{ m}$) & $M_{2}$ ($M_{\odot}$)  &   $d(R_{1},M_{1},R_{2},M_{2})$ \\
\hline
$1.35$ & $1.2$ & $1.4$ & $1.2$ & $5.8 \times 10^{5}$ \\
$1.11$ & $1.2$ & $1.16$ & $1.2$ & $6.0 \times 10^{5}$ \\
$1.0$ & $1.2$ & $1.04$ & $1.2$ & $6.2 \times 10^{5}$ \\
\hline
$1.2$ & $1.2$ & $1.2$ & $1.3$ & $1.5 \times 10^{4}$ \\
$1.2$ & $1.3$ & $1.2$ & $1.4$ & $1.6 \times 10^{4}$ \\
$1.2$ & $1.4$ & $1.2$ & $2.0$ & $1.3 \times 10^{5}$ \\
\hline
\hline
\end{tabular}
\label{tab:simdata}
\end{table*}

\subsection{Geodesic paths}

To further explore the structure of the configuration space spanned by Tolman VII stars, we investigate geodesic paths. While it is not clear if these curves have any physical relevance beyond being used to measure distance through \eqref{eq:distance}, it seems plausible that least action principles, applied to the lengths of curves within $\text{Met}_{\text{NS}}(M)$, might imply something about stellar evolution.

To this end, the problem may be thought about as follows: consider a star initially in some state, $(R_{1},M_{1})$, evolving towards a different state, $(R_{2},M_{2})$, through some physical process. Suppose that, whatever this process may be, the star evolves so as to minimise an energy integral on some appropriate configuration space, which may (or may not) be the space $\text{Met}_{\text{NS}}(M)$. As is well-known, geodesics, which extremise arc-length, also extremise energy \cite{carmo}, and therefore trace some kind of energy-minimising evolution. Again, whether this is relevant to stellar dynamics is unclear, though, in any case, it is interesting to explore the mathematical structure of the configuration manifold.

\begin{figure*}
\includegraphics[width=0.493\textwidth]{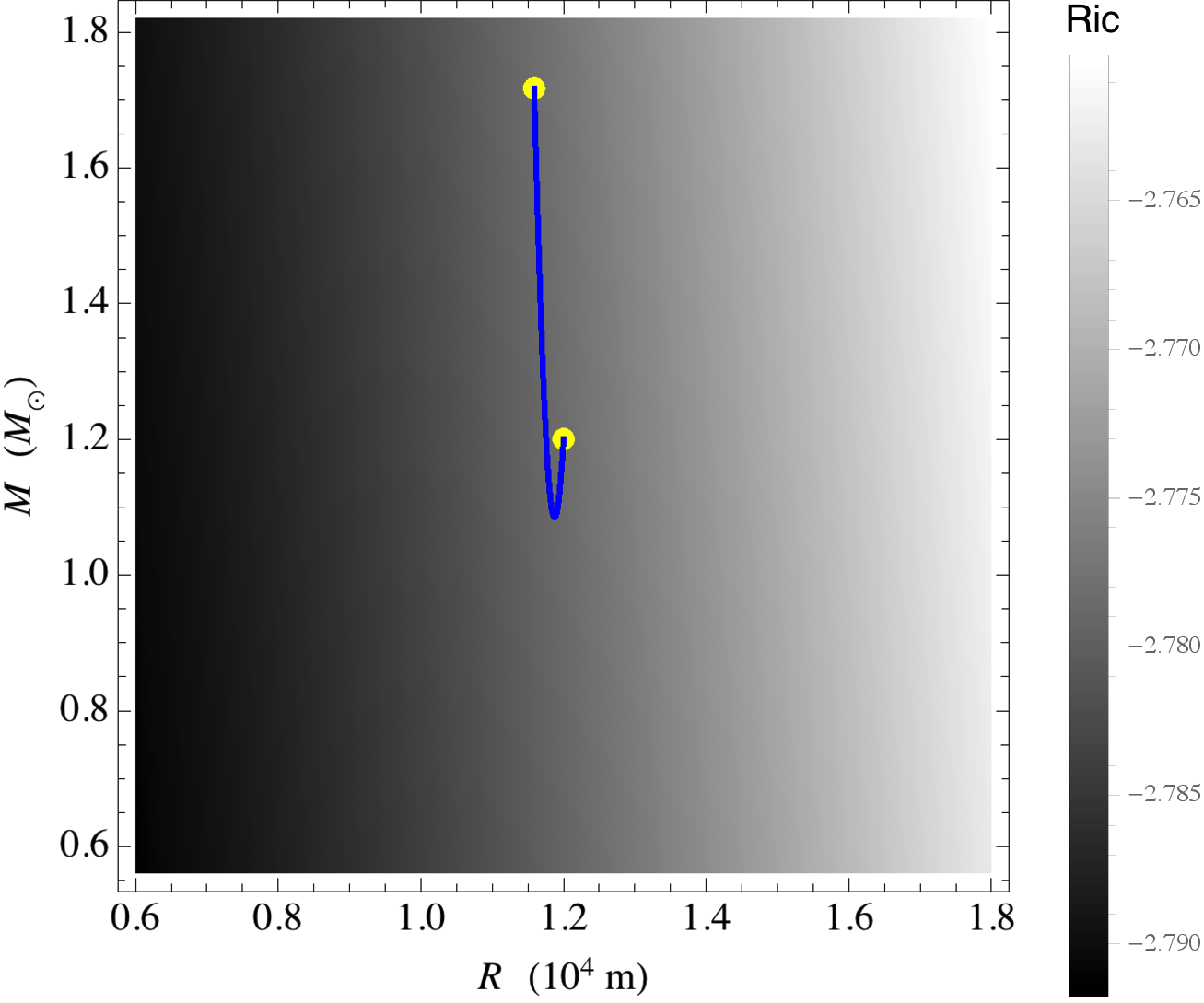}
\caption{Geodesic path (blue curve) on the configuration manifold of Tolman VII metrics, connecting the points (yellow dots) $(R,M)=(1.2 \times 10^{4} \text{ m}, 1.2 M_{\odot})$ and $(R,M)=(1.15 \times 10^{4} \text{ m}, 1.75 M_{\odot})$. The colour scale shows the (suitably normalised) Ricci scalar $G_{i j} R_{G}^{i j}$, with darker shades indicating a greater value for $|G_{i j} R_{G}^{i j}|$.  \label{fig:fig1}}
\end{figure*}

Figure \ref{fig:fig1} presents the geodesic curve on the Tolman VII configuration manifold connecting the points $(R,M)=(1.2 \times 10^{4} \text{ m}, 1.2 M_{\odot})$ and $(R,M)=(1.15 \times 10^{4} \text{ m}, 1.75 M_{\odot})$, with the (suitably normalised) Ricci scalar curvature $\text{Ric} = R_{G}^{ij} G_{ij}$ of $\text{Met}_{\text{T-VII}}(M)$ with $\bar{R} = 1.8 \times 10^{4} \text{ m}$. Loosely speaking, the scalar curvature traces how the volume form deviates from its flat counterpart, and thus affects how length is measured. We see that the geodesic path connecting the end points exhibits significant curvature, indicating that the configuration manifold has a complicated geometric structure. The curve further suggests that a star evolving, from the initial to the final states defined by the end points of the geodesic, may have non-monotonic behaviour in the relative mass and radius shifts which occur during the state change. 



\section{Discussion}

In this paper we explore a mathematical framework to quantify the `distance' between different NS models. In particular, many different stellar models have been proposed in the literature \cite{nseos1,nseos2,most18}, some members of which should be, intuitively speaking, `closer' together than others. The framework developed here allows for a rigorous definition of `closeness', by defining a distance, given by expression \eqref{eq:distance}, on the configuration space of NS models, $\text{Met}_{\text{NS}}(M)$. We have shown how the framework may be applied in the simple case of Tolman VII stars, and have speculated that geodesics on this configuration manifold may imply something about stellar evolution beyond providing a distance measure; see Fig. \ref{fig:fig1}. While the work presented here is mostly conceptual, it is hoped that it may be useful in future studies of NS structure.

It is interesting to note that, as for the initial considerations by Wheeler and others \cite{wheeler1,wheeler2,guil09}, nothing within the formalism developed here explicitly restricts us to NS spacetimes. For example, an extension to black hole spacetimes could be developed, though there are certain obstacles. In particular, the construction of the space $M$ from $S$ is not obvious in this case, since the asymptotic behaviour of the black hole may be relevant, e.g. asymptotically de Sitter black holes behave differently to asymptotically flat ones \cite{dsbh}, and a distance measure should reflect this. This is problematic since the compactness of $M$, which cannot be imposed if one wishes to integrate out to infinity, is assumed so that \eqref{eq:gilmed} is well-defined. If some compact hypersurface $M \subset \mathcal{M}$ can be constructed in an invariant manner which captures the black hole physics, or if suitably decaying conformal factors can be introduced so that \eqref{eq:gilmed} converges {(i.e. build $h_{ij}$ from $e^{2 \Omega} \kappa_{\mu \nu}$ with conformal factor $\Omega$ decaying sufficiently rapidly so that \eqref{eq:gilmed} converges)} \cite{hansen}, the formalism developed here would largely carry over. This could be used to quantify the `closeness' of black hole models in different modified theories of gravity \cite{bhmod1,bhmod2,bhmod3}.

\acknowledgements
We thank Prof. Bill Moran for introducing us to several key references. {We thank the anonymous referees for their useful feedback.} This work was supported by the Alexander von Humboldt Foundation.

\end{document}